\documentclass[twocolumn,showpacs,amsfonts,aps,prc,floatfix,%
superscriptaddress,nofootinbib]{revtex4}

\usepackage{amsmath}
\usepackage{bm}
\usepackage{graphicx}

\newcommand{\Tdec}{T_\mathrm{dec}}

\newcommand{\Tchem}{T_\mathrm{chem}}

\newcommand{\VC}{{\tt VISHNU}}

\newcommand{\ecc}{\varepsilon}
\newcommand{\dNdeta}{dN_\mathrm{ch}/d\eta}
\newcommand{\dNdy}{dN_\mathrm{ch}/dy}
\newcommand{\Npart}{N_\mathrm{part}}

\newcommand{\be}[1]{\begin{equation}\label{#1}}
\newcommand{\ee}{\end{equation}}

\newcommand{\eq}{{\,=\,}}

\def\La{\langle}
\def\Ra{\rangle}

\usepackage{color}

\begin{document}


\title{Radial and elliptic flow in Pb+Pb collisions at the Large Hadron 
Collider\\ from viscous hydrodynamics}

\author{Chun Shen}
\email[Correspond to\ ]{shen@mps.ohio-state.edu}
\affiliation{Department of Physics, The Ohio State University, 
  Columbus, OH 43210-1117, USA}
\author{Ulrich Heinz}
\email[Email:\ ]{heinz@mps.ohio-state.edu}
\affiliation{Department of Physics, The Ohio State University, 
  Columbus, OH 43210-1117, USA}
\author{Pasi Huovinen}
\email[Email:\ ]{huovinen@th.physik.uni-frankfurt.de}
\affiliation{Institut f\"ur Theoretische Physik, Johann Wolfgang 
Goethe-Universit\"at, Max-von-Laue-Stra\ss e 1, D-60438 Frankfurt 
am Main, Germany}
\author{Huichao Song}
\email[Email:\ ]{HSong@LBL.gov}
\affiliation{Lawrence Berkeley National Laboratory, 1 Cyclotron Road,
             Berkeley, CA 94720, USA}
\date{\today}

\begin{abstract}
A comprehensive viscous hydrodynamic fit of spectra and elliptic flow for 
charged hadrons and identified pions and protons from Au+Au collisions of 
all centralities measured at the Relativistic Heavy Ion Collider is 
performed and used as the basis for predicting the analogous observables 
for Pb+Pb collisions at the Large Hadron Collider at $\sqrt{s}\eq2.76$ 
and 5.5\,$A$\,TeV. Comparison with recent measurements of the elliptic flow 
of charged hadrons by the ALICE experiment shows that the model slightly 
over-predicts the data if the same (constant) specific shear viscosity 
$\eta/s$ is assumed at both collision energies. In spite of differences in 
our assumptions for the equation of state, the freeze-out temperature, the 
chemical composition at freeze-out, and the starting time for the 
hydrodynamic evolution, our results agree remarkably well with those
of Luzum [M. Luzum, Phys.\ Rev.\  C {\bf 83}, 044911 (2011)], indicating 
robustness of the hydrodynamic model extrapolations. Future measurements 
of the centrality and transverse momentum dependence of spectra and elliptic 
flow for identified hadrons predicted here will further test the model and 
shed light on possible variations of the quark-gluon transport coefficients
between RHIC and LHC energies.   
\end{abstract}

\maketitle

\section{Introduction}
\label{sec1}

The first measurement of elliptic flow in Pb+Pb collisions at the 
Large Hadron Collider (LHC) has just been reported \cite{Aamodt:2010pa}.
The elliptic flow coefficient $v_2$ characterizes the momentum anisotropy 
of final particle emission in non-central heavy-ion collisions relative
to the event-plane, defined by the beam direction and the minor axis
of the nuclear overlap region in the collision. It describes the efficiency
of the medium generated in the collision to generate from an initial spatial
deformation of its density distribution an asymmetry in the final momentum
distribution, through interactions of the medium constituents. This
efficiency increases with the coupling strength between those constituents
and becomes maximal for an infinitely strongly coupled medium. In this
limit the mean free path of the constituents becomes as small as allowed by
the uncertainty relation \cite{Danielewicz:1984ww}, and the medium reaches 
very quickly a state of approximate local thermal equilibrium which
allows to describe its evolution with fluid dynamics. For given
initial spatial deformation of the collision fireball, ideal fluid
dynamics (which assumes zero mean free path) is expected to generate 
the largest possible elliptic flow \cite{Heinz:2001xi}. Shear viscosity, a 
consequence of non-zero mean free paths and limited from below by quantum 
mechanics \cite{Danielewicz:1984ww,Policastro:2001yc}, will lead to a 
suppression of $v_2$ \cite{Heinz:2002rs,Teaney:2003kp}.

Compelling evidence for fluid dynamical behavior of the collision 
fireballs created in ultrarelativistic heavy-ion collisions, with a
very small ratio of shear viscosity to entropy density, $\eta/s$, 
has been found in heavy-ion collisions at the Relativistic Heavy Ion 
Collider (RHIC) \cite{Arsene:2004fa,Lacey:2006pn,Romatschke:2007mq,%
Song:2010mg}. The new data from the LHC confirm this picture
\cite{Aamodt:2010pa,Luzum:2010ag,Lacey:2010ej} and agree, at least 
qualitatively, with hydrodynamic predictions of elliptic flow for Pb+Pb 
collisions at the LHC \cite{Abreu:2007kv,Niemi:2008ta,Kestin:2008bh,%
Luzum:2009sb,Bozek:2010wt,Hirano:2010jg,Schenke:2011tv}.

The purpose of the present article is to explore how good this agreement
is quantitatively, and to what extent the present and future LHC elliptic
flow data can tell us novel facts about the transport behavior of hot
QCD matter at temperatures that exceed those accessible at RHIC but are 
within reach at the LHC. Similar to the analyses in \cite{Niemi:2008ta,%
Luzum:2009sb,Luzum:2010ag,Schenke:2011tv}, but different from recent 
hybrid model studies in \cite{Hirano:2010jg,Song:2011qa}, we base our 
analysis on a purely hydrodynamic approach. While this ignores the fact 
that the late dilute hadronic stage of the expansion is very dissipative 
and not well described by fluid dynamics (neither ideal \cite{Hirano:2005xf} 
nor viscous \cite{Song:2010aq}), the importance of the hadronic phase for 
the development of elliptic flow is expected to be much reduced at the LHC 
relative to RHIC \cite{Hirano:2007xd,Hirano:2010jg}. As in
Refs.~\cite{Luzum:2009sb,Luzum:2010ag,Schenke:2011tv,Song:2011qa}, but 
different from Refs.~\cite{Niemi:2008ta,Hirano:2010jg}, we use {\em viscous} 
hydrodynamics with a non-zero (but constant, i.e. temperature independent) 
specific shear viscosity $\eta/s$, adjusted to spectra and elliptic flow 
measurements at RHIC. Our fitted value $\eta/s\eq0.20$ (for CGC initial 
conditions, see below) is 25\% larger than that used by Luzum and Romatschke 
\cite{Luzum:2009sb,Luzum:2010ag} but agrees well with the value for the 
quark-gluon plasma (QGP) viscosity $(\eta/s)_\mathrm{QGP}$ recently extracted 
from RHIC data by using a hybrid viscous hydrodynamic + Boltzmann approach 
(\VC\ \cite{Song:2010mg,Song:2010aq}). Calculations with such a hybrid approach 
are numerically much more demanding than purely hydrodynamic simulations; 
a generalization of the present analysis using \VC\ will follow soon and 
should further improve the reliability of the LHC predictions.  

\section{Hydrodynamic fit of RHIC Au+Au data}
\label{sec2}

In this work, we employ (2+1)-d viscous hydrodynamics \cite{Song:2007fn} 
with the lattice QCD based equation of state {\tt s95p-PCE} 
\cite{Huovinen:2009yb,Shen:2010uy}, which accounts for chemical freeze-out 
before thermal decoupling at $\Tchem\eq165$\,MeV, to simulate
the expansion of the collision 
fireball. From the analysis \cite{Song:2010mg} of charged hadron spectra 
and elliptic flow in $200\,A$\,GeV Au+Au collisions at RHIC we take over a 
value of $\eta/s\eq0.20$ (corresponding to MC-KLN initial conditions, see 
below) for the effective specific shear viscosity of the strongly interacting 
fluid. Using the insights obtained from the systematic parameter study 
presented in \cite{Shen:2010uy}, we initialize the hydrodynamic expansion at 
time $\tau_0 = 0.6$\,fm/$c$ and decouple at $\Tdec\eq120$\,MeV at both RHIC 
and LHC energies. For Au+Au collisions at RHIC energies these parameters allow 
for a good global description of the hadron $p_T$-spectra and differential 
elliptic flow (see below). Lacking strong theoretical or phenomenological 
guidance how to adjust their values for Pb+Pb collisions at the LHC, we 
here decided to keep them unchanged. 

At thermalization time $\tau_0$, we assume that the shear stress
tensor is given by its Navier-Stokes value $\pi^{\mu\nu}\eq2\eta
\sigma^{\mu\nu}$ (where 
$\sigma^{\mu\nu}\eq{\nabla^{\left\La\mu\right.}\!}{u^{\!\left.\nu\right\Ra}}$ 
is the symmetric and traceless velocity shear tensor),
and that the initial expansion flow is entirely longitudinal with Bjorken 
profile $v_z\eq{z/t}$ and zero transverse flow velocity. In Milne 
coordinates $(\tau,x,y,\eta)$ this corresponds to an initial flow
4-velocity $(u^\tau,u^x,u^y,u^\eta)\eq(1,0,0,0)$. Kinetic freeze-out
is implemented by converting the hydrodynamic output to particle
spectra with the Cooper-Frye prescription \cite{Cooper:1974mv}
on a decoupling surface of constant temperature $\Tdec$. We use
a quadratic ansatz $\delta f(x,p){\,\sim\,}p^\mu p^\nu\pi_{\mu\nu}(x)$
\cite{Teaney:2003kp} for the viscous deviation from local thermal 
equilibrium of the local phase-space distribution function on the 
freeze-out surface. Our final hadron spectra include decay products 
from strong decays of all particles and resonances up to 2\,GeV mass 
\cite{Sollfrank:1990qz}, using the resonance decay program from the 
{\tt AZHYDRO} package.\footnote{{\tt AZHYDRO} is available at the URL\\ 
   {\tt http://www.physics.ohio-state.edu/\~{}froderma/}.}

For the initial density profile we here use a Monte-Carlo version 
\cite{Drescher:2006pi,Drescher:2007ax} of the Kharzeev-Levin-Nardi 
\cite{KLN} model (MC-KLN).\footnote{The Monte Carlo code is 
  available at URL \\ {\tt http://www.aiu.ac.jp/\~{}ynara/mckln/}.}
The specific implementation used in this work is described in 
\cite{Hirano:2009ah,Hirano:2010jg}. The model gives for each event the 
gluon density distribution in the transverse plane. We assume it to 
thermalize by time $\tau_0$, and convert the gluon density into entropy 
density \cite{Hirano:2005xf}. Over one million Monte Carlo events are 
re-centered to the beam axis and rotated in the transverse plane such 
that their minor axis aligns with the impact parameter (i.e. their 
``participant plane'' coincides with the reaction plane). After sorting 
them into collision centrality bins according to their number $\Npart$ of 
participating (``wounded'') nucleons, we average them to obtain a smooth 
average initial entropy density which is then converted to energy 
density using the equation of state. Using this smooth energy density 
as weight, we compute the initial eccentricity 
$\bar{\ecc}\eq\frac{\La y^2{-}x^2\Ra}{\La y^2{+}x^2\Ra}$ and overlap area 
$S\eq\pi\sqrt{\La x^2\Ra\La y^2\Ra}$ of the reaction zone; these represent 
the corresponding mean values of events in this centrality 
class.\footnote{Note that about 10\% larger overlap areas are
   obtained when using the entropy density as weight
   \cite{Hirano:2010jg,Song:2010mg}, whereas for all but the most 
   central collisions the eccentricities of the energy and entropy 
   densities are nearly identical \cite{Qiu:2011iv}.}

\begin{figure}[ht]
\includegraphics[width=0.95\linewidth,clip=]{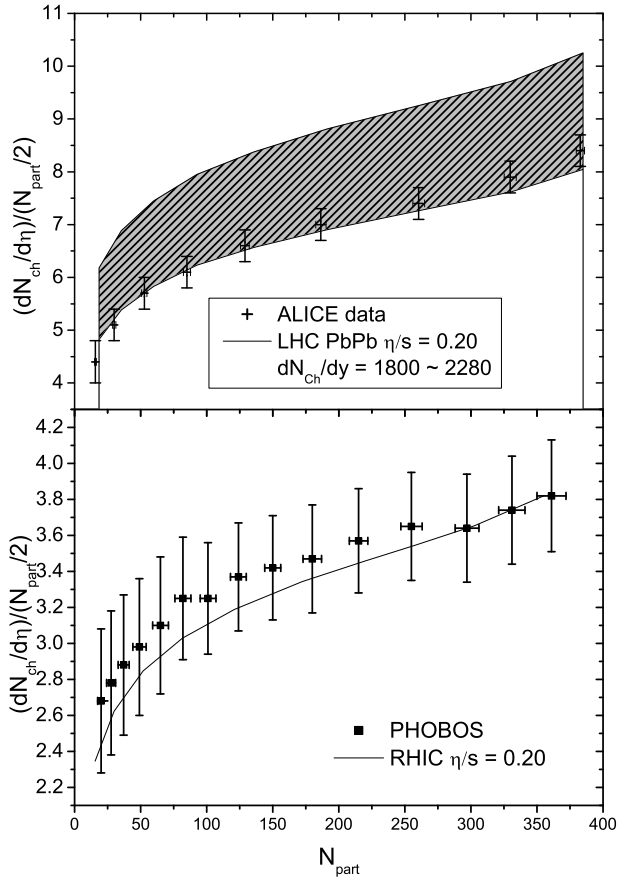}
\caption{Centrality dependence of the charged hadron multiplicity per unit 
pseudo-rapidity, $\dNdeta/(\Npart/2)$ as a function of $\Npart$, in 
$200\,A$\,GeV Au+Au collisions at RHIC (bottom panel) and in 
$(2.76-5.5)\,A$\,TeV Pb+Pb collisions at the LHC (top panel). Experimental 
data are from the PHOBOS Collaboration \cite{Back:2004dy} for Au+Au 
collisions at $\sqrt{s}\eq200\,A$\,GeV, and from the ALICE Collaboration 
\cite{Aamodt:2010cz} for Pb+Pb collisions at $\sqrt{s}\eq2.76\,A$\,TeV. 
The lines are from the MC-KLN model (see text). For Au+Au at RHIC the 
MC-KLN model was normalized to the measured multiplicity in the 0-5\% 
centrality bin; at the LHC, the lines bounding the shaded region were 
normalized to $\dNdeta\eq1548$ and 1972 (or $\dNdy\eq1800$ and 2280) 
at 0-5\% centrality, respectively.
\label{F1} 
}
\end{figure}

The KLN model involves a couple of parameters that need to be adjusted
to obtain the correct final charged hadron multiplicity $\dNdeta$ in 
central Au+Au collisions at RHIC. In \cite{Hirano:2009ah} this adjustment
was performed for ideal fluid dynamics (which conserves entropy) coupled
to a hadron cascade. The model then correctly predicts the charged hadron
multiplicities at all other collision centralities. In our viscous 
hydrodynamic model, viscous heating produces additional entropy, leading 
to somewhat larger final multiplicities. We thus perform an iterative
renormalization of the initial entropy density profile until the measured 
charged hadron multiplicity in the $0{-}5\%$ most central $200\,A$\,GeV 
Au+Au collisions at RHIC is once again reproduced. The lower panel of 
Fig.~\ref{F1} shows that, after this renormalization, the model again 
correctly describes the measured \cite{Back:2004dy} centrality dependence 
of charged hadron production. The centrality dependence of viscous entropy 
production (which is relatively larger in peripheral than in central 
collisions \cite{Song:2007fn}) is (at least at RHIC energies) sufficiently 
weak to not destroy the agreement of the model with experimental
observations. 

The ability of the MC-KLN model to describe the centrality dependence of 
charged hadron production without additional parameters is the main reason 
for choosing it over the MC-Glauber model as our basis for extrapolation 
from RHIC to LHC energies. It was recently shown \cite{ALbacete:2010ad}
that this centrality dependence is robust against running coupling 
corrections \cite{Balitsky:2006wa,Albacete:2007sm,Kovchegov:2006vj} in 
the Balitsky-Kovchegov evolution (on which the KLN model is based) which
were found to hardly affect its shape. They do, however, modify the 
collision energy dependence of particle production, with the LHC Pb+Pb 
data being better described if running coupling corrections are included 
\cite{Aamodt:2010cz}. Our version of the MC-KLN model does not include 
running coupling corrections,\footnote{The {\tt rcBK} code in
\cite{ALbacete:2010ad} 
includes running coupling corrections but it has not been renormalized 
to take into account viscous entropy production at RHIC energies.}
and we must normalize the initial entropy density profile 
for Pb+Pb collisions at the LHC separately from Au+Au collisions at RHIC.
Without such an independent renormalization, we overpredict the measured 
charged multiplicity from central $2.76\,A$\,TeV Pb+Pb collisions 
\cite{Aamodt:2010pb,Aamodt:2010cz} by about 10\%. 

After renormalization we obtain the solid lines bounding the shaded
region in the upper panel of Fig.~\ref{F1}, with the lower (upper)
bound corresponding to Pb+Pb collisions at 2.76 (5.5) $A$\,GeV,
respectively. The data in that panel are from the ALICE Collaboration
for Pb+Pb at $2.76\,A$\,GeV \cite{Aamodt:2010pb,Aamodt:2010cz}. (For 
$5.5\,A$\,GeV Pb+Pb collisions we assumed $\dNdy\eq2280$ (corresponding
to $\dNdeta\eq1972$), based on an extrapolation of Fig.~3 in 
Ref.~\cite{Aamodt:2010pb}.) One sees that, even without running coupling
corrections, but including viscous entropy production, the MC-KLN model
does a good job in describing the measured centrality dependence of
charged hadron production in Pb+Pb collisions at the LHC. This gives hope
that the successful description of the centrality dependence of hadron
spectra and elliptic flow at RHIC energies (see below and \cite{Song:2010mg})
translates into a reliable prediction of the corresponding centrality
dependences in Pb+Pb collisions at the LHC.
  
\begin{figure}[h!]
\includegraphics[width=\linewidth,clip=]{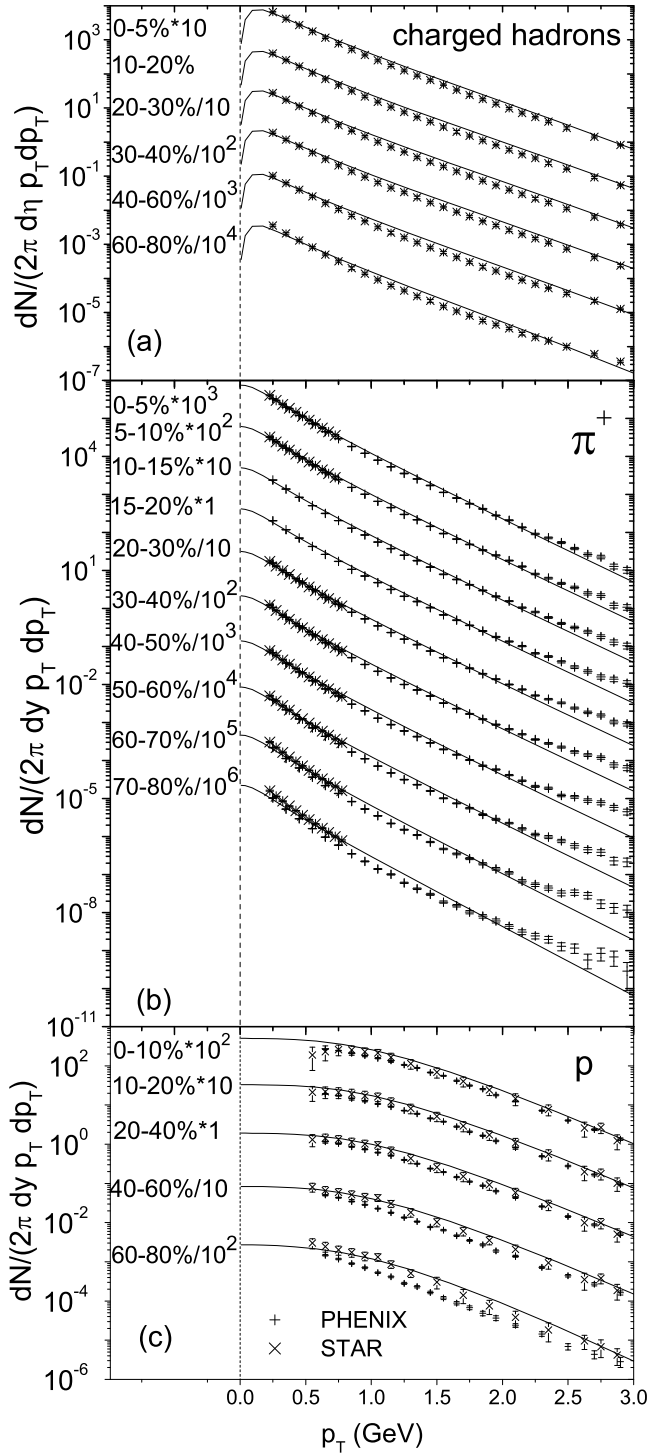}
\caption{\label{F2} $p_T$-spectra of all charged hadrons (a), positive pions
(b) and protons (c) for 200\,$A$\,GeV Au+Au collisions of different 
centralities as indicated. The symbols show data from the STAR 
(\cite{Adams:2003kv,Adams:2003xp,:2008ez}, $\times$) and PHENIX 
(\cite{Adler:2003au,Adler:2003cb}, $+$) experiments. The lines are 
results from the viscous hydrodynamic model for constant $\eta/s\eq0.20$ 
and MC-KLN initial conditions (see text for other model parameters).   
}
\end{figure}

\begin{figure}[t]
\includegraphics[width=\linewidth,clip=]{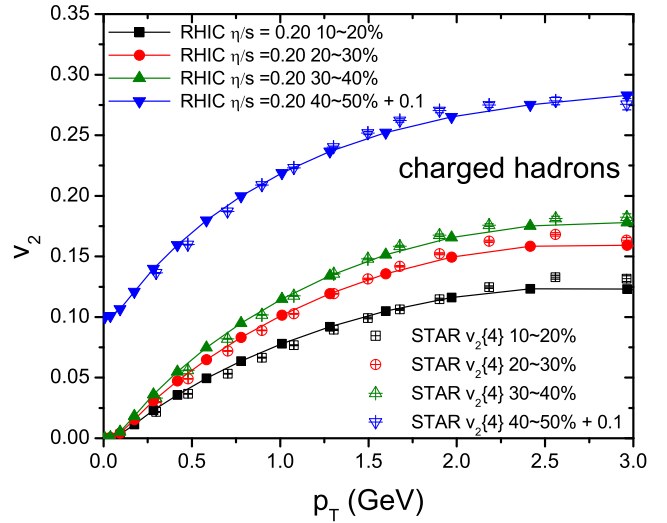}
\caption{(Color online) Differential elliptic flow $v_2(p_T)$
for charged hadrons from 200\,$A$\,GeV Au+Au collisions of different 
centralities as indicated. Open symbols are experimental data from the
STAR experiment for $v_2\{4\}(p_T)$ \cite{BaiThesis,:2008ed}, lines 
with filled symbols of the same shape are the corresponding 
hydrodynamic fits with the same model as in Fig.~\ref{F1}. For the 
$40{-}50\%$ centrality bin data and theory are vertically offset by 0.1 
for better visibility.
\label{F3} 
}
\end{figure}

Figures~\ref{F2} and \ref{F3} establish our baseline for the extrapolation
to LHC energies. In Fig. \ref{F2} we show our purely hydrodynamic fit 
(obtained with parameters $\tau_0$, $\eta/s$, $\Tchem$, and $\Tdec$ set as described 
above\footnote{Note that our value $\tau_0\eq0.6$\,fm/$c$ is 45\% smaller
  than the value of 1.05\,fm/$c$ used for $\eta/s\eq0.2$ in the \VC\ 
  simulations in \cite{Song:2010mg}. The earlier evolution of hydrodynamic 
  transverse flow arising from this smaller $\tau_0$ value compensates for 
  the lack of a highly dissipative hadronic phase in the purely hydrodynamic 
  approach. Hadronic dissipation leads to a significant broadening in 
  particular of the proton $p_T$-spectra during the hadronic 
  stage which (given the constraints from the elliptic flow data which
  prohibit us from simply lowering $\Tdec$) viscous hydrodynamics with 
  temperature-independent $\eta/s\eq0.2$ cannot replicate.})
of the hadron spectra measured in $200\,A$\,GeV Au+Au collisions at 
RHIC. Fig.~\ref{F2}a shows the mid-rapidity transverse momentum spectra 
per unit pseudo-rapidity for unidentified charged hadrons from the STAR
\cite{Adams:2003kv} and PHENIX \cite{Adler:2003au} experiments compared 
with the hydrodynamical model. Figs.~\ref{F2}b,c show a similar comparison 
for the $p_T$-spectra per unit rapidity of identified pions and protons 
from STAR \cite{Adams:2003xp,:2008ez} and PHENIX \cite{Adler:2003cb}.
In the experimental spectra, protons from weak decays were removed;
STAR quotes a large systematic error associated with this feeddown
correction, and within that large error band the two data sets agree
with each other, even if the central values of the STAR proton data
appear to be up to 50\% higher than PHENIX data. Our results agree well 
with the STAR protons for $p_T > 0.6$\,GeV/$c$ but overpredict the PHENIX 
protons by up to a factor 2. 

Figure~\ref{F3} shows the hydrodynamically calculated differential elliptic 
flow for unidentified charged hadrons in comparison with STAR $v_2\{4\}(p_T)$ 
data \cite{BaiThesis,:2008ed}, for four centrality classes ranging from 
semi-central to mid-peripheral collisions ($10{-}50\%$ centrality). With 
$\eta/s = 0.20$, viscous hydrodynamics gives an excellent description 
of the STAR $v_2\{4\}$ data, even up to 3\,GeV/$c$ in transverse 
momentum (i.e. beyond the $p_T$ range where the hydrodynamic description
is expected to begin to break down, due to the increasing influence of
hard production processes and large uncertainties in the viscous correction
$\delta f$ to the local phase-space distribution at kinetic freeze-out 
\cite{Shen:2011kn}). Looking more carefully one sees that our model 
slightly overestimates the elliptic flow at low $p_T{\,<\,}1$\,GeV while 
underestimating it in the high-$p_T$ region, $p_T{\,>\,}2$\,GeV. 

In \cite{Shen:2010uy} we noted a tension in trying to simultaneously 
fit within a purely viscous hydrodynamic approach the proton $p_T$-spectra 
and the charged hadron differential elliptic $v^{\mathrm{ch}}_2\{2\}(p_T)$ 
when using EOS {\tt s95p-PCE}. Even a temperature dependent $\eta/s(T)$ 
that has a larger shear viscosity in the hadronic phase could not resolve 
this tension: in \cite{Shen:2011kn} two of us found that
the RHIC Au+Au hadron 
spectra are insensitive to a temperature-dependent increase of the shear 
viscosity in the hadron gas phase, as was previously seen in 
\cite{Niemi:2011ix}. Figs.~\ref{F2} and \ref{F3} demonstrate that this 
problem is largely resolved when using the $v_2\{4\}(p_T)$ data 
(Fig.~\ref{F3}) instead of $v_2\{2\}$ (see Fig.~\ref{F8} further below): 
We obtain an excellent description of the differential elliptic flow, 
together with an acceptable description (within large experimental 
uncertainties) of the $p_T$ spectra. 

Overall, the viscous fluid dynamic description of the hadron spectra 
and charged hadron elliptic flow $v_2(p_T)$ shown here is of similar 
quality as the hybrid model description with \VC\ presented in 
\cite{Song:2010mg}. Since purely hydrodynamic simulations are numerically 
much less costly than calculations with \VC, we will now use them to 
generate a broad range of predictions for soft hadron production in Pb+Pb 
collisions at the LHC.   

\section{Predictions for Pb+Pb collisions at the LHC}
\label{sec3}

As discussed above, the extrapolation from RHIC to LHC is done keeping
$\tau_0,\,\Tchem,\,\Tdec$ and $\eta/s$ fixed. When comparing the resulting
viscous hydrodynamic predictions with experimental data from the recently
started LHC heavy-ion collision program, we will search for indications
from experiment that would motivate changing these parameters. First
results for $p_T$-spectra \cite{Aamodt:2010jd} as well as both the
$p_T$-differential and $p_T$-integrated elliptic flow of unidentified 
charged hadrons \cite{Aamodt:2010pa} have already been published and
will be compared with the theoretical predictions below. Additional 
experimental information on spectra and elliptic flow of identified 
hadrons will become available soon; the relevant hydrodynamic predictions
are presented in this section.  

\begin{figure}[h!]
\includegraphics[width=0.9\linewidth,clip=]{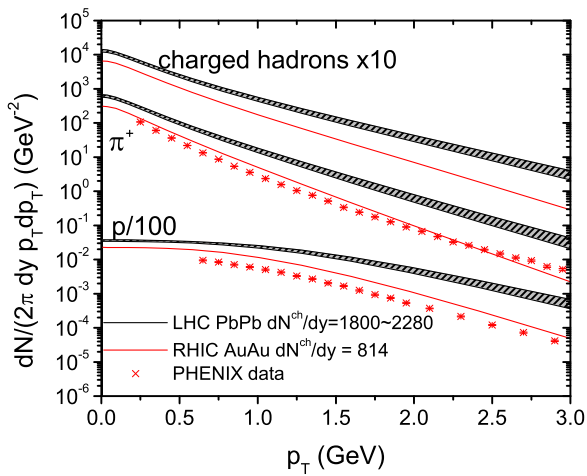}
\caption{(Color online) $p_T$-spectra of all charged hadrons, positive pions, 
and protons for minimum bias 200\,$A$\,GeV Au+Au (thin red lines and 
data points) and $(2.76{-}5.5)\,A$\,TeV Pb+Pb collisions (black lines
with shaded area). The RHIC data are from the PHENIX experiment
\cite{Adler:2003cb}. The shaded bands for the LHC predictions are limited
at the bottom (top) by lines for $\sqrt{s}\eq2.76\ (5.5)\ A$\,TeV, 
corresponding to $\dNdy\eq1800$ (2280) 
($dN_\mathrm{ch}/d\eta\eq1548\ (1972)$). The calculations assume the 
same constant $\eta/s\eq0.2$ at all shown collision energies. 
\label{F4}
}
\end{figure}

In Fig.~\ref{F4} we show the transverse momentum spectra for all charged 
hadrons, as well as for identified pions and protons, for minimum bias
collisions of Au+Au at RHIC and Pb+Pb at the LHC.\footnote{To simulate 
  minimum bias collisions, we compute the spectra for the centrality 
  classes shown in Figs.~\ref{F2}(b) and \ref{F6} and average them. Any 
  additional observables, such as the minimum bias elliptic flow in 
  Fig.~\ref{F8} below, are calculated from these minimum bias spectra.} 
For RHIC we compare with data from the PHENIX Collaboration 
\cite{Adler:2003cb}. The upper and lower bounds of the shaded areas are 
predictions for minimum bias Pb+Pb collisions at collision energies of 
5.5 and 2.76\,TeV per nucleon pair, respectively. The LHC spectra are 
visibly flatter than at RHIC energies, reflecting stronger radial flow. 
For central collisions ($0{-}5\%$ centrality), the fireball lifetime 
increases from Au+Au at RHIC to Pb+Pb at LHC by about 19\% and 24\%, 
respectively, for 2.76 and $5.5\,A$\,TeV collision energy; for 
peripheral collisions at $70{-}80\%$ centrality, the corresponding 
lifetime increases are even larger (34\% and 41\%, respectively). The 
average radial flow velocity increases in central collisions ($0{-}5\%$ 
centrality) by 5 and 7\%, respectively, and in peripheral collisions 
($70{-}80\%$ centrality) by 9 and 11\%.

\begin{figure}[h!]
\includegraphics[width=\linewidth,clip=]{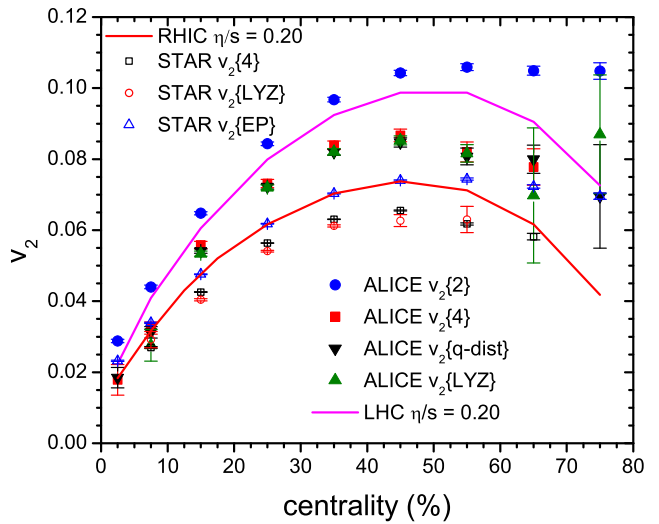}
\caption{(Color online) $p_T$-integrated elliptic flow of charged hadrons
for 200\,$A$\,GeV Au+Au collisions at RHIC (open symbols are STAR data 
\cite{BaiThesis,:2008ed}, the lower red line is the result from
viscous hydrodynamics) and for 2.76\,$A$\,TeV Pb+Pb collisions at the
LHC (filled symbols are ALICE data \cite{Aamodt:2010pa}, the upper 
magenta line shows the viscous hydrodynamic prediction). In both experiment
and theory the differential elliptic flow $v_2(p_T)$ (see Figs.~\ref{F3}
and \ref{F7}) was integrated over the range 
0.15\,GeV/$c{\,<\,}p_T{\,<\,}2$\,GeV/$c$ for Au+Au at RHIC and over
$p_T{\,>\,}0.2$\,GeV/$c$ for Pb+Pb at the LHC. 
\label{F5}
}
\end{figure}

\begin{figure*}
\hspace*{-2mm}
\includegraphics[width=0.9\linewidth,clip=]{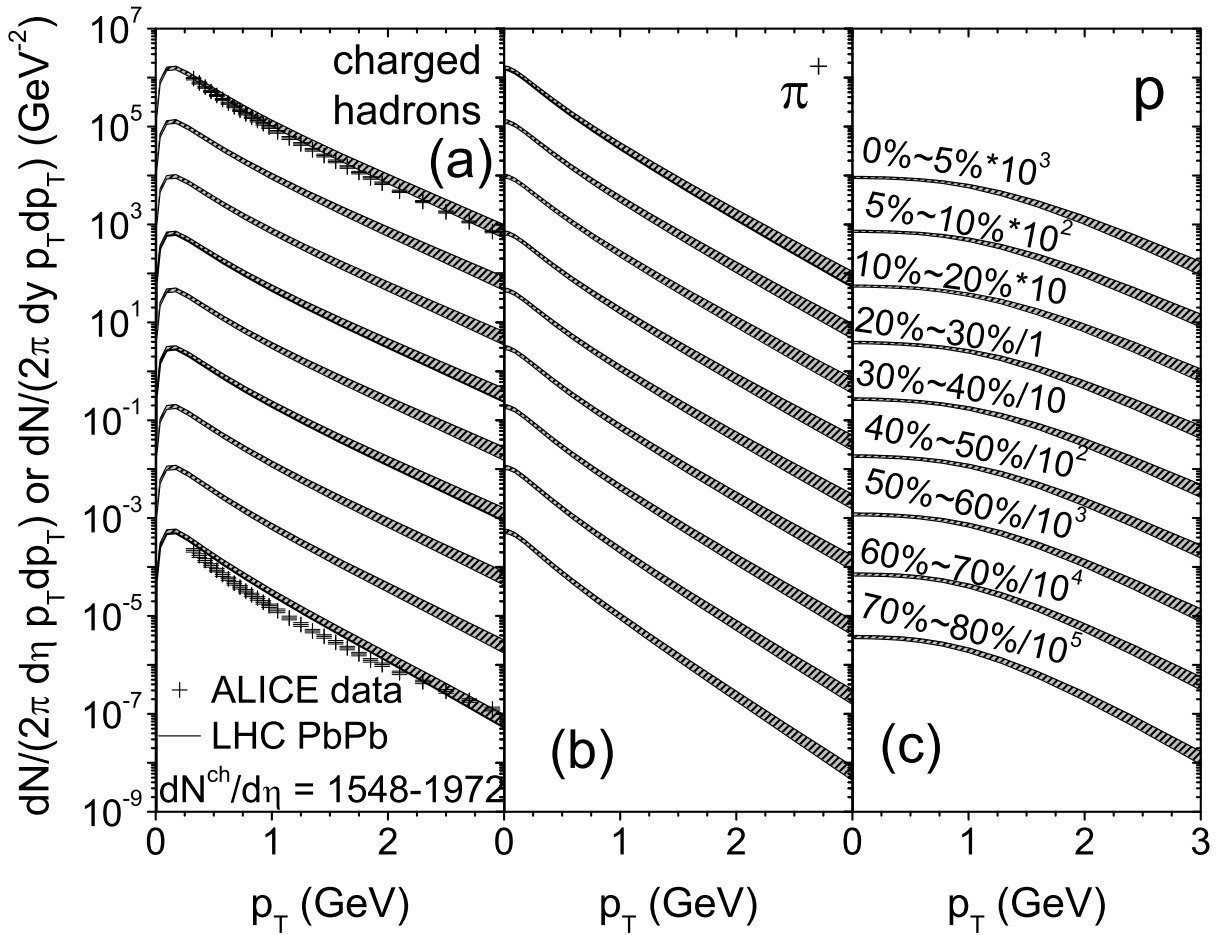}
\caption{\label{F6} $p_T$-spectra per unit pseudorapidity for charged
  hadrons (a) and per unit rapidity for pions (b) and protons (c) for Pb+Pb 
  collisions at the LHC. The lower and upper end of the shaded bands 
  represent viscous hydrodynamic predictions for $\sqrt{s}\eq2.76$ and 
  $5.5\,A$\,TeV (corresponding to $dN_\mathrm{ch}/d\eta\eq1548$ and 1972, 
  or $\dNdy\eq1800$ and 2280), respectively. Experimental data in panel (a)
  are from the ALICE experiment \cite{Aamodt:2010jd}.
}
\end{figure*}

Figure \ref{F5} shows the integrated charged hadron elliptic flow $v_2$ 
as a function of collision centrality for Au+Au collisions at RHIC
and Pb+Pb collisions at the LHC. At RHIC energy, our results (lower red 
line) overestimates the STAR $v_2\{4\}$ data by about 11\% in 
mid-central collisions, but agrees nicely with  $v_2\{\mathrm{EP}\}$
except for the most peripheral collisions.\footnote{In very peripheral 
  collisions, the fireball lifetime decreases dramatically, cutting short 
  the buildup of anisotropic hydrodynamic flow and thereby prohibiting 
  $v_2$ from saturating. In addition, viscous effects are stronger in the 
  small fireballs created in peripheral collisions than in the larger central
  collision fireballs. Both effects together cause the theoretical
  $v_2$ values to decrease sharply at large collision centralities,
  in apparent conflict with the experimental data. The experimental 
  $v_2\{2\}$ and $v_2\{\mathrm{EP}\}$ measurements are, however, 
  contaminated by non-flow effects, in particular in very peripheral 
  collisions. Once non-flow effects are corrected for 
  \cite{:2011vk}, the experimental $v_2$ values decrease
  at large collision centralities much in the same way as predicted 
  by hydrodynamics.}
At first sight the overprediction of the $p_T$-integrated $v_2\{4\}$ at 
RHIC is surprising, given the excellent description of the differential
elliptic flow $v_2\{4\}(p_T)$ shown in Fig.~\ref{F3}. The apparent paradox
is resolved by observing that the hydrodnamically computed charged hadron
$p_T$-spectra shown in Fig.~\ref{F2} are somewhat harder than measured,
thereby giving too much weight in the $p_T$-integral to the range 
$0.75 < p_T < 2$\,GeV/$c$ where $v_2\{4\}(p_T)$ is large.\footnote{The 
   agreement with the $v_2\{\mathrm{EP}\}$ data is fortuitous and should, 
   in fact, not happen since the measured $v_2\{\mathrm{EP}\}$ includes a 
   positive contribution from event-by-event $v_2$ fluctuations 
   \cite{Ollitrault:2009ie} while our hydrodynamic calculation yields 
   the average elliptic flow $\La v_2\Ra$ which is smaller.}

At LHC energy ($\sqrt{s}\eq2.76\,A$\,TeV) our integrated $v_2$ lies 
between $v_2\{2\}$ and $v_2\{4\}$ values measured by the ALICE Collaboration 
\cite{Aamodt:2010pa}. Again, we overpredict the $p_T$-integrated $v_2\{4\}$ 
by about $10{-}15\%$. We note that from RHIC to LHC the hydrodynamically 
computed integrated 
$v_2$ in mid-central to mid-peripheral collisions increases by about 
30\%, in agreement with the experimental observations. This is due to 
reduced viscous suppression effects in the larger and denser fireballs
created at the LHC and a longer fireball lifetime which allows the 
momentum flow anisotropy to approach saturation more closely than at
lower energies \cite{Kolb:2003dz,Hirano:2007xd}. In very peripheral 
collisions, even at LHC energies such a saturation of $v_2$ does not 
happen; this is the reason why in Fig.~\ref{F5} the integrated $v_2$ 
is seen to decrease at large collision centralities, both at RHIC and 
LHC.  

\begin{figure*}
\hspace*{-2mm}
\includegraphics[width=0.9\linewidth,clip=]{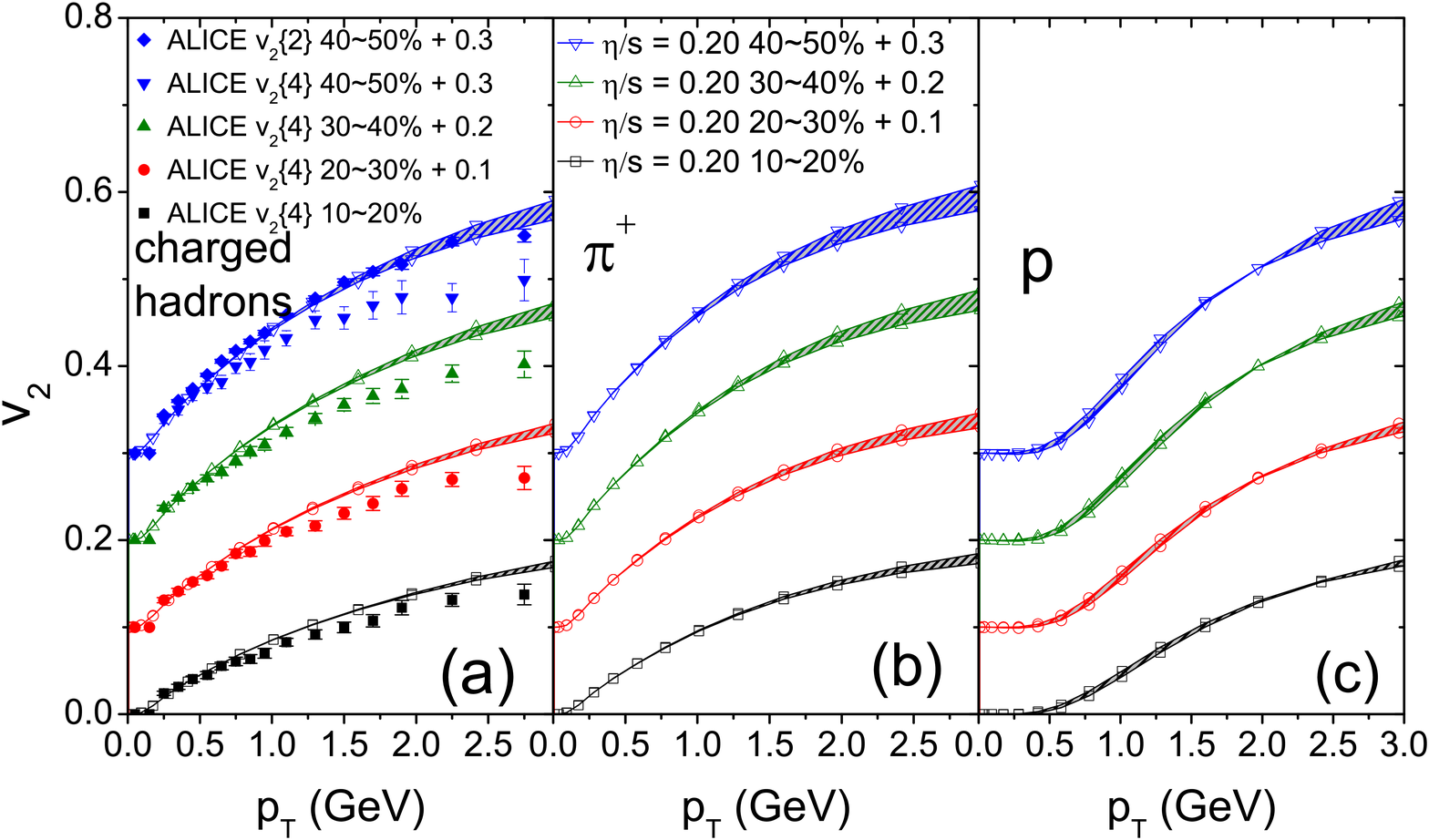}
\caption{(Color online) Differential elliptic flow $v_2(p_T)$ for 
  unidentified charged hadrons (a) and identified pions (b) and protons (c), 
  for Pb+Pb collisions of different centralities at the LHC. Experimental
  data for charged hadron $v_2(p_T)$, denoted by solid symbols, are from 
  the ALICE experiment \cite{Aamodt:2010pa}; they should be compared with 
  theoretical lines carrying open symbols of the same shape and color.
  The shaded bands show the variation of the hydrodynamic predictions 
  with collision energy between $\sqrt{s}\eq2.76$ and 5.5\,TeV (corresponding 
  to $\dNdy\eq1800$ and 2280, respectively). The lines corresponding 
  to the lower collision energy ($\sqrt{s}\eq2.76$\,TeV) define the lower 
  end of the shaded regions at $p_T\eq3$\,GeV/$c$. 
\label{F7} 
}
\end{figure*}

In Fig.~\ref{F6} we present hadron transverse momentum spectra for Pb+Pb 
collisions at LHC energies, for a range of collision centralities. In
panel (a) we compare the hydrodynamic predictions with first data
from the ALICE experiment \cite{Aamodt:2010jd}. Overall, the 
theoretical description of these experimental data is of similar quality 
as for Au+Au collisions at RHIC (see Fig.~\ref{F2}). In the most central
collisions, the hydrodynamical model describes the charged hadron spectrum
somewhat better than at RHIC, whereas in the very peripheral collisions 
the hydrodynamic spectra are too flat, presumably due to large viscous shear
pressure effects. Future comparison with the measured spectra at other
collision centralities and for identified hadrons, shown here in panels (b) 
and (c) as predictions, should shed further light on the origin of the
discrepancy in peripheral collisions.

Figure~\ref{F7}a shows a comparison of the hydrodynamically generated 
differential $v_2^\mathrm{ch}(p_T)$ for charged hadrons with the ALICE 
$v_2\{4\}$ data \cite{Aamodt:2010pa}, for four different collision 
centralities. For the most peripheral of these, we also show the measured
$v_2\{2\}$ for comparison. The hydrodynamic predictions agree nicely
with the data at low $p_T<1$\,GeV/$c$, but overshoot the experimental 
values by $10{-}20\%$ at larger $p_T$, especially in the more peripheral
bins. In the $40{-}50\%$ centrality bin, the theoretical prediction 
happens to agree nicely with $v_2\{2\}(p_T)$ even though the latter 
should be shifted upward by flow fluctuations that are not included in the 
theoretical calculation. We note that the theoretical overshoot is less
severe in the {\tt VISHNU} hybrid model (see Fig.~3 in \cite{Song:2011qa}) 
than in the purely hydrodynamic simulations shown here. This suggests
that the excess of $v_2(p_T)$ over the measured values at $p_T>1$\,GeV/$c$ 
in Fig.~\ref{F7}a may be caused by an inadequate description of the late 
hadronic stage and its freeze-out. 

We can summarize Figs.~\ref{F2}a, \ref{F3}, \ref{F5}, \ref{F6}, 
and \ref{F7}a by observing that the hydrodynamic model overpredicts 
the $p_T$-integrated charged hadron $v_2$ by $10{-}15\%$ at both 
RHIC and LHC, but for different reasons: at RHIC the differential
elliptic flow $v_{2,\mathrm{ch}}(p_T)$ is correctly reproduced while 
the inverse slope of the theoretical $p_T$-spectra is slightly too 
large, while the LHC $p_T$-spectra are described a bit better (at least 
in the most central collisions where published data are available) but 
the slope of $v_{2,\mathrm{ch}}(p_T)$ at the LHC is slightly 
overpredicted.

Panels (b) and (c) of Fig.~\ref{F7} give predictions for the differential 
$v_2(p_T)$ of identified pions and protons. Please note the different 
shape of the proton $v_2(p_T)$ from that of the pions at low $p_T$: radial
flow pushes the proton elliptic flow to larger values of $p_T$. Comparing
the curves for $\sqrt{s}\eq2.76$ and $5.5\,A$\,TeV, we see that this
``radial push'' of the proton $v_2$ increases with collision energy,
so for higher $\sqrt{s}$ the rise of $v_2(p_T)$ is shifted to larger
transverse momenta, while at fixed $p_T<1.5$\,GeV/$c$ the proton elliptic
flow {\em decreases} with increasing collision energy. This happens only
for heavy hadrons but not for the much lighter pions (see panel (b)).
  
\begin{figure}[h!]
\includegraphics[width=\linewidth,clip=]{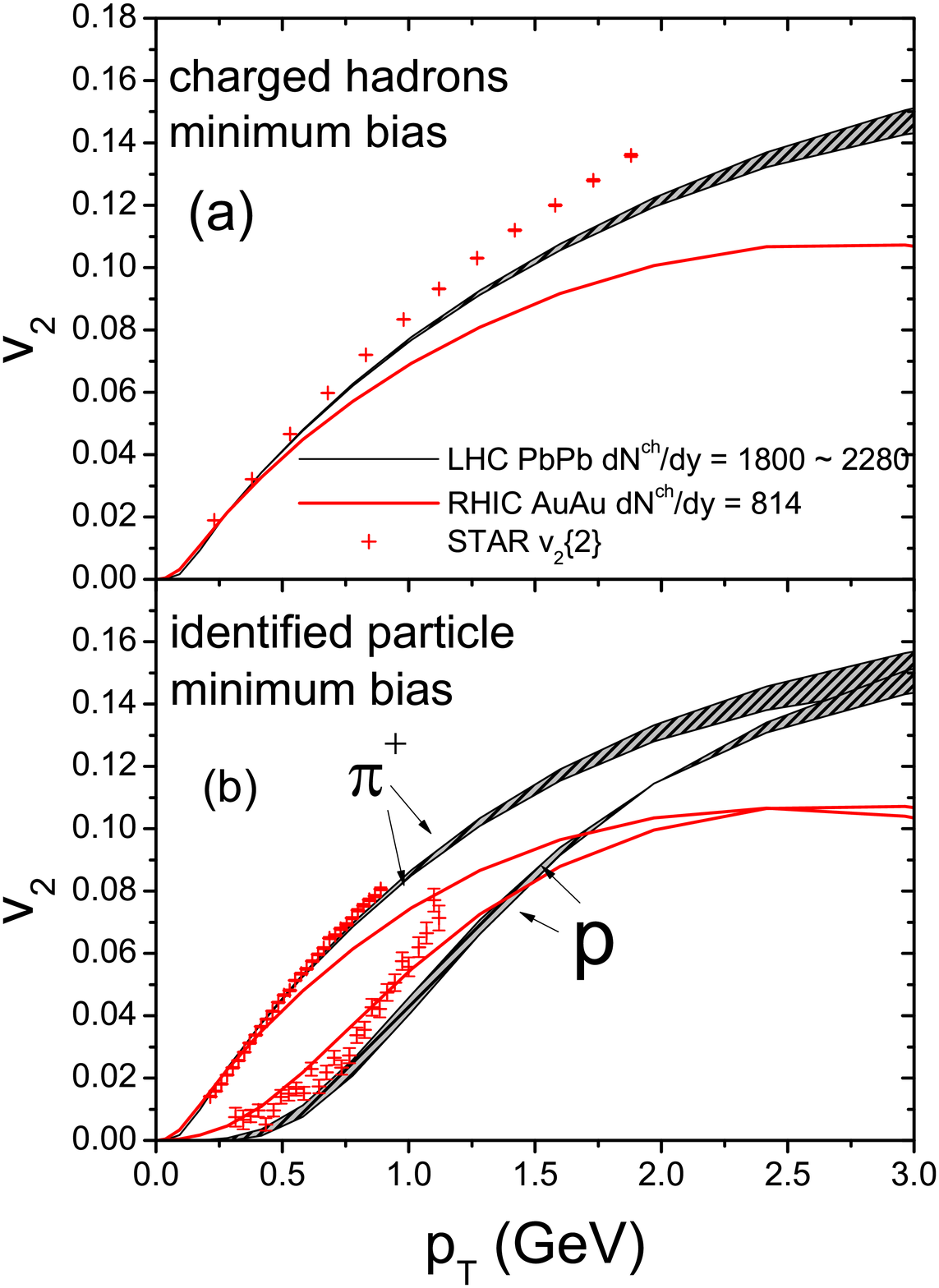}
\caption{(Color online) Differential elliptic flow $v_2(p_T)$ for all 
  charged hadrons (a) and identified pions and protons (b), for minimum
  bias 200\,$A$\,GeV Au+Au collisions at RHIC and $(2.76-5.5)\,A$\,TeV 
  Pb+Pb collisions at the LHC. Experimental data for $v_2\{2\}$ from Au+Au 
  collisions at RHIC are from the STAR experiment \cite{Adams:2004bi}.
  Solid lines are viscous hydrodynamic results for 200\,$A$\,GeV Au+Au 
  collisions with the same hydrodynamic parameters as in 
  Figs.~\ref{F1}-\ref{F4}; note their disagreement with the $v_2\{2\}$ 
  data shown here (in contrast to their excellent agreement with $v_2\{4\}$ 
  data shown in Fig.~\ref{F3}).
  The shaded bands are LHC predictions and show the variation of the  
  theoretical expectations for Pb+Pb collisions at collision energies 
  ranging from $\sqrt{s}\eq2.76$ to $5.5\,A$\,TeV (corresponding to 
  $\dNdy\eq1800$ and 2280, respectively). As in Fig.~\ref{F7}, the 
  lines defining the lower end of the shaded region at $p_T\eq3$\,GeV/$c$
  correspond to the lower LHC energy $\sqrt{s}\eq2.76\,A$\,TeV.
\label{F8}
}
\end{figure}

\begin{figure*}
\includegraphics[width=0.82\linewidth,clip=]{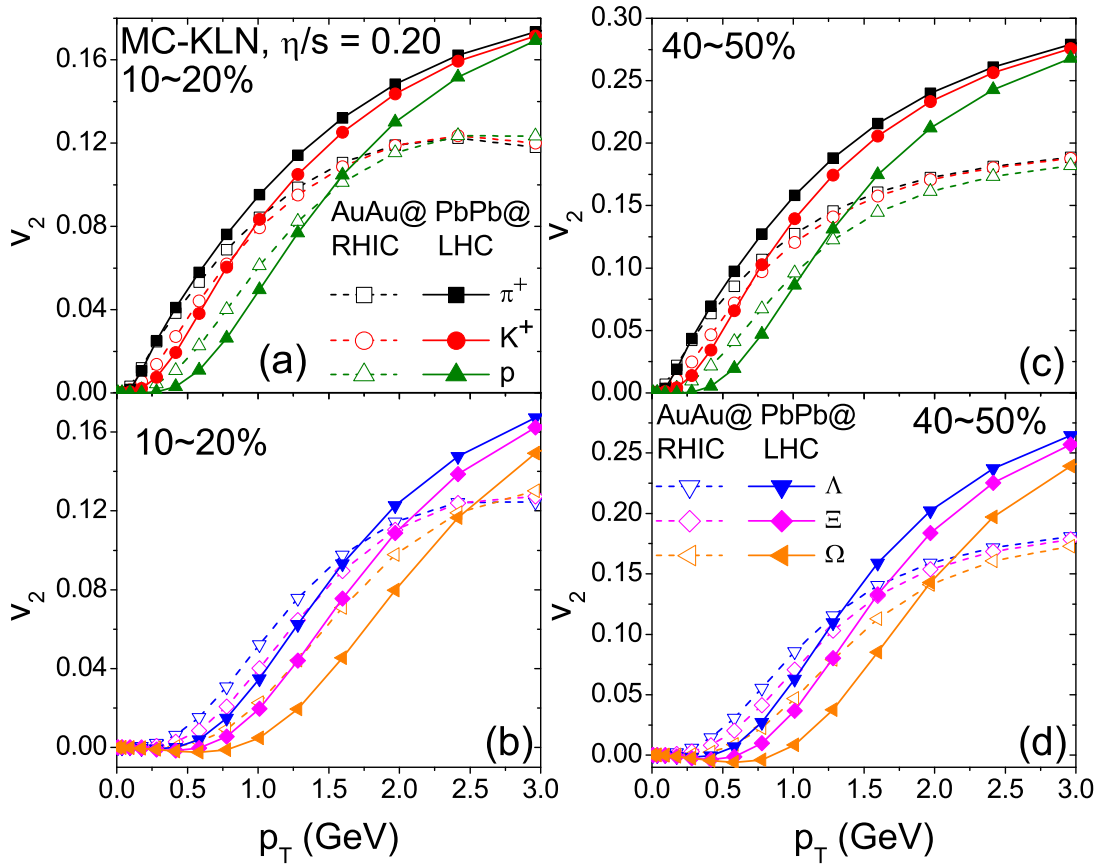}
\caption{(Color online) Comparison of the differential elliptic flow 
  $v_2(p_T)$ for $200\,A$\,GeV Au+Au collisions at RHIC (dashed lines)
  and $2.76\,A$\,TeV Pb+Pb collisions at the LHC (solid lines), at 
  $10\%{-}20\%$ (a,b) and $40\%{-}50\%$ (c,d) centrality, for a 
  variety of different hadron species. Note the slightly negative 
  elliptic flow for the heavy $\Omega$ hyperons at low $p_T$.
\label{F9}
}
\end{figure*}

In Figs.~\ref{F8} and \ref{F9} we pursue this theme further, by
directly comparing the differential elliptic flows at RHIC and LHC 
energies. In Fig.~\ref{F8} we show results for minimum bias collisions; 
the RHIC predictions are compared with available data from STAR 
\cite{Adams:2004bi}. We see that at low $p_T$, the elliptic flow
for unidentified charged hadrons (which are strongly pion dominated) and
for identified pions increases from RHIC to LHC whereas the opposite is 
true for protons. At higher $p_T$ ($p_T > 1.5$\,GeV/$c$), on the other
hand, $v_2(p_T)$ increases for {\em both} pions and protons as we increase
the collision energy. Fig.~\ref{F9} shows this for a few more hadron
species, for the $10{-}20\%$ and $40{-}50\%$ centrality bins: the heavier 
the hadron, the stronger a push of $v_2$ towards higher $p_T$ is observed. 
At sufficiently large $p_T$, $v_2(p_T)$ is larger at LHC than at RHIC for 
all particle species, but at low $p_T$ this holds only for pions whereas
all heavier hadrons show a decrease of $v_2(p_T)$ from RHIC to LHC at
fixed $p_T$. As the hadron rest mass grows, the crossing point where the 
decrease of $v_2$ at fixed $p_T$ with rising collision energy turns into 
an increase shifts to larger $p_T$ values. In view of Fig.~\ref{F9}, the 
experimental observation \cite{Aamodt:2010pa} that for unidentified charged 
hadrons $v_2^\mathrm{ch}(p_T)$ hardly changes at all from RHIC to LHC appears
accidental:\footnote{Contrary to the claim made in \cite{Lacey:2010ej},
  the observation that the ratio between $v_2^\mathrm{ch}(p_T)$ measured at
  LHC and at RHIC is approximately independent of $p_T$ cannot be directly
  used to conclude that $(\eta/s)_\mathrm{QGP}$ does not change from RHIC 
  to LHC. If that argument were correct, this ratio should be independent
  of $p_T$ not only for the sum of all charged hadrons, but also for each
  identified hadron species separately. Our hydrodynamic calculations show 
  that the latter does not hold even if $\eta/s$ remains unchanged from 
  RHIC to LHC.} 
The increase of $v_2(p_T)$ at fixed $p_T$ for pions is balanced by a 
corresponding decrease for all heavier hadrons leaving, as it happens, 
no visible net effect once all charged hadrons are lumped together.

\begin{figure*}
\includegraphics[width=0.95\linewidth,clip=]{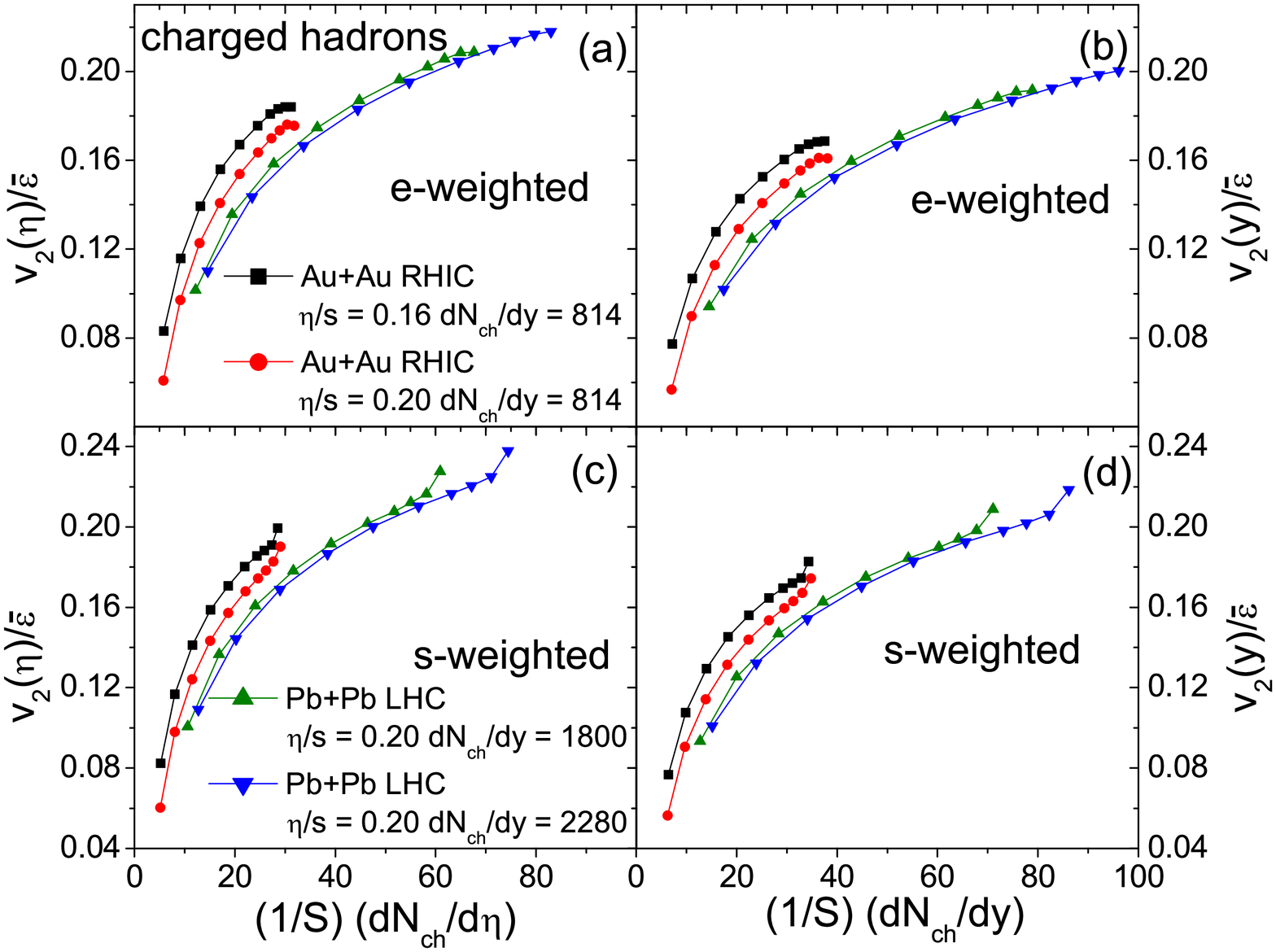}
\caption{(Color online) Eccentricity-scaled elliptic flow 
  $v_2/\bar{\ecc}$ as function of the charged hadron multiplicity density
  per unit overlap area $S$ from viscous hydrodynamic calculations at 
  $\sqrt{s}\eq0.2,\,2.76,$ and 5.5\,$A$\,TeV (corresponding to 
  $\dNdy\eq814$ (Au+Au), 1800 and 2280 (Pb+Pb), respectively). Each line 
  corresponds to one collision system at fixed collision energy but 
  different collision centralities (from right to 
  left, the symbols correspond to 0-5\%, 5-10\%, 10-15\%, 15-20\%,
  20-30\%, 30-40\%, 40-50\%, 50-60\%, 60-70\% and 70-80\% centrality). 
  The four panels show $v_2(\eta)/\bar{\ecc}$ vs. $(1/S)(\dNdeta)$ 
  (where $\eta$ is pseudorapidity) (a,c) and $v_2(y)/\bar{\ecc}$
  vs. $(1/S)(\dNdy)$ (where $y$ is rapidity) (b,d), with $\bar{\ecc}$ and
  $S$ evaluated with the participant-plane averaged energy density 
  $\bar{e}(\bm{r}_\perp,\tau_0)$ as weight function (a,b) (default 
  option, see Sec.~\ref{sec2}) or (for comparison with other work) with 
  the corresponding entropy density $\bar{s}(\bm{r}_\perp,\tau_0)$ as 
  weight (c,d). The RHIC curves for $\eta/s\eq0.16$ (black squares) 
  illustrate the effect of changing the value of the specific shear 
  viscosity by 0.04. The LHC calculations are done for $\eta/s\eq0.20$ 
  as obtained from the hydrodynamic fit to RHIC data.
\label{F10}  
}
\end{figure*}

In Refs.~\cite{Song:2010mg} it was argued that a robust method for 
extracting the QGP shear viscosity is to fit the collision centrality 
dependence of the eccentricity-scaled charged hadron elliptic flow 
$v_2^\mathrm{ch}/\bar{\ecc}$ with a viscous hydrodynamic + hadron cascade
hybrid code. In that study it was found that, at fixed collision
energy,\footnote{We recently checked that this
    multiplicity scaling carries over to other collision systems such
    as Cu+Cu {\em at the same collision energy} \cite{SSH}.}
plotting $v_2^\mathrm{ch}/\bar{\ecc}$ against the charged hadron
multiplicity density per unit overlap area, $(1/S)(\dNdy)$, yields 
``universal'' curves that depend only on the QGP shear viscosity but
not on the model for the initial energy density distribution (in 
particular its eccentricity). In Fig.~\ref{F10} we show such a plot
for $200\,A$\,GeV Au+Au collisions at RHIC together with Pb+Pb collisions
at two LHC energies. The four panels show this scaling in terms 
of distributions in pseudo-rapidity ($\eta$, left column) or rapidity 
($y$, right column), and also compare it for our default choice of using 
the initial {\em energy} density as weight for the calculation of the 
average eccentricity $\bar{\ecc}$ and overlap area $S$ (top row) 
with what one obtains by evaluating these quantities with the initial 
{\em entropy} density instead (as is done in 
Refs.~\cite{Hirano:2010jg,Song:2010mg}) (bottom row). We see that,
independent of these choices of representation, the universality of 
$v_2^\mathrm{ch}/\bar{\ecc}$ vs. $(1/S)(\dNdy)$ or $(1/S)(\dNdeta)$ does 
not carry over to different collision energies (at least not for the 
purely hydrodynamic simulations studied in the present work):
At the same multiplicity density $(1/S)(\dNdy)$ or $(1/S)(\dNdeta)$, 
more peripheral higher energy collisions produce less elliptic flow 
per initial eccentricity than more central lower energy collisions. At 
fixed $\eta/s\eq0.2$, the difference between $200\,A$\,GeV Au+Au and 
$2.76\,A$\,TeV Pb+Pb collisions (red circles vs. green upward triangles in 
Fig.~\ref{F10}) is as large as the difference between $\eta/s\eq0.16$ 
and $\eta/s\eq0.20$ for Au+Au collisions at fixed $\sqrt{s}\eq200\,A$\,GeV 
(red circles vs. black squares). 

We note that the tendency in Fig.~\ref{F10} of higher energy
collisions producing less $v_2^\mathrm{ch}/\bar{\ecc}$ at fixed
$(1/S)(\dNdy)$ than lower energy ones contradicts the opposite
tendency observed by Hirano {\it et al.} in Fig.~3 of
Ref.~\cite{Hirano:2010jg} where an ideal hydro + hadron cascade hybrid
code was employed.\footnote{The careful reader will notice that for
  200\,$A$\,GeV Au+Au collisions, our maximal values for
  $(1/S)(\dNdeta)$ in Fig.~\ref{F10}b are significantly larger than
  those shown in Fig.~3 of Ref.~\cite{Hirano:2010jg}. This is due to a
  lower normalization of the initial entropy density in
  \cite{Hirano:2010jg}, corresponding to $\dNdeta\sim600$ instead of
  our $\dNdeta\sim700$ in central Au+Au collisions (T. Hirano, private
  communication).}  The authors of \cite{Hirano:2010jg} presented
strong arguments that their observation of larger
$v_2^\mathrm{ch}/\bar{\ecc}$ at fixed $(1/S)(\dNdeta)$ in higher
energy collisions is not related to their use of a hadron cascade for
describing the late hadronic stage. Our opposite finding, on the other
hand, is supported by the earlier purely
hydrodynamic scaling studies presented in the last two works of
\cite{Song:2007fn} whose authors came to the same
conclusion as we do here. At present this discrepancy remains
unresolved; we suspect, however, that the origin of the difference
between our work and that of Hirano {\it et al.} could
  be in their use of a more realistic (3+1)-dimensional hydrodynamic
evolution~\cite{Monnai:2011ju}, 
although in the earlier ideal fluid hydrodynamical
  studies at the full RHIC energy, the differences between boost invariant and
  non-boost invariant results were small~\cite{Heinz:2001xi,Hirano:2002ds}.
Possible consequences of
the violation of boost-invariance in RHIC and LHC heavy-ion collisions
are presently being studied \cite{SSH}.

Before moving on, let us comment on the different shape at the 
high-multiplicity end of the curves shown 
in Fig.~\ref{F10} when using entropy instead of energy density
as the weight for calculating the initial eccentricity $\bar{\ecc}$
overlap area $S$: It is caused by the different centrality dependence
of the energy and entropy density weighted eccentricities in near-central
collisions observed in Ref.~\cite{Qiu:2011iv} whose authors showed that 
in the most central collisions (where $\bar{\ecc}$ is dominated by 
event-by-event shape fluctuations) the entropy-weighted participant 
eccentricity decreases faster with decreasing impact parameter than the   
energy-weighted one.

\section{Temperature dependent $\bm{\eta/s(T)}$}
\label{sec4}

Shear viscosity is known to suppress the buildup of elliptic flow.
Naively, the systematic overprediction of $v_2\{4\}(p_T)$ in Pb+Pb 
collisions at the LHC seen in Fig.~\ref{F7}a, together with the 
excellent description of the same quantity in Au+Au collisions at RHIC
seen in Fig.~\ref{F3}, thus suggests that the fireball liquid formed
in LHC collisions might be slightly more viscous (i.e. possess larger
average $\eta/s$) than at RHIC energies \cite{Niemi:2011ix,Song:2011qa}. 
In this section we present some results using a temperature dependent
specific shear viscosity, $(\frac{\eta}{s})(T)$, that were motivated by
such considerations.

\begin{figure}[ht]
\includegraphics[width=\linewidth,clip=]{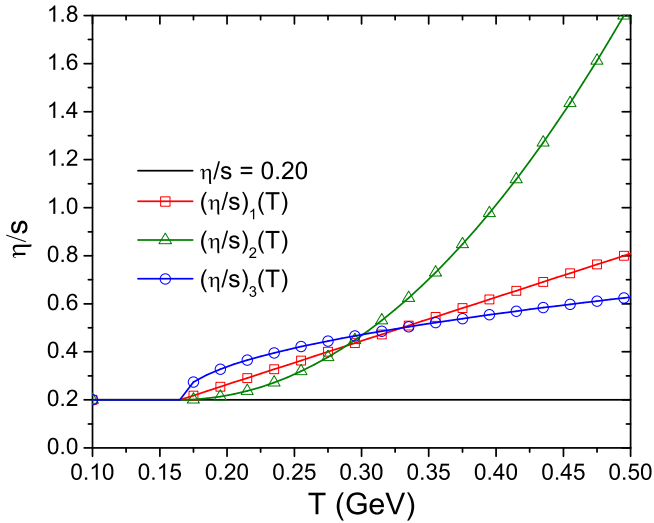}
\caption{(Color online) Three temperature dependent parametrizations
$(\eta/s)(T)$ studied in this section. In all cases, $\eta/s\eq0.2$ for
$T{\,<\,}\Tchem\eq165$\,MeV.
\label{F11} 
}
\end{figure}

Figure~\ref{F11} illustrates the following three trial functions 
explored in this section:
\begin{eqnarray}
\label{eq1}
  \left(\frac{\eta}{s}\right)_1 &=& 0.2 + 0.3\,\frac{T{-}\Tchem}{\Tchem},
\\
\label{eq2}
  \left(\frac{\eta}{s}\right)_2 &=& 0.2 + 0.4\,\frac{(T{-}\Tchem)^2}
  {T^2_\mathrm{chem}},
\\
\label{eq3}
  \left(\frac{\eta}{s}\right)_3 &=& 0.2 + 0.3\, 
  \sqrt{\frac{T{-}\Tchem}{\Tchem}}.
\end{eqnarray}
Here $\Tchem\eq165$\,MeV is the chemical decoupling temperature and stands
for the ``transition temperature'' at which the hadronization of quarks
and gluons is complete.

In principle, the value of $\eta/s$ should exhibit a minimum near 
$\Tchem$ and increase again in the hadronic phase below $\Tchem$
\cite{Csernai:2006zz,Chen:2007jq,Kapusta:2008vb}. The authors of 
\cite{Niemi:2011ix} pointed out, however, that at the full LHC 
collision energy of 5.5\,$A$\,TeV the behavior of $\eta/s$ at temperatures 
below $\Tchem$ has very little effect on the final hadron spectra and 
their elliptic flow. At 2.76\,$A$\,TeV the effect on elliptic flow was 
moderate, and negligible on the spectra. Here we will concentrate on 
qualitative aspects of effects arising from a temperature dependent 
growths of $\eta/s$ in the high temperature region that can be explored 
at LHC energies but is beyond the reach of RHIC, and continue to set 
$\eta/s\eq0.2$ at $T{\,<\,}\Tchem$ for simplicity.

As pointed out in \cite{Niemi:2011ix}, the spectra and elliptic flow 
in Au+Au collisions at RHIC energies are most sensitive to the average
value of $\eta/s$ in the temperature region below 220-230\,MeV. We have 
checked that altering $\eta/s$ at higher temperatures as shown in 
Fig.~\ref{F11} has little influence on the results at RHIC energies
shown in Sec.~\ref{sec2}.

\begin{figure}[htb]
\includegraphics[width=\linewidth,clip=]{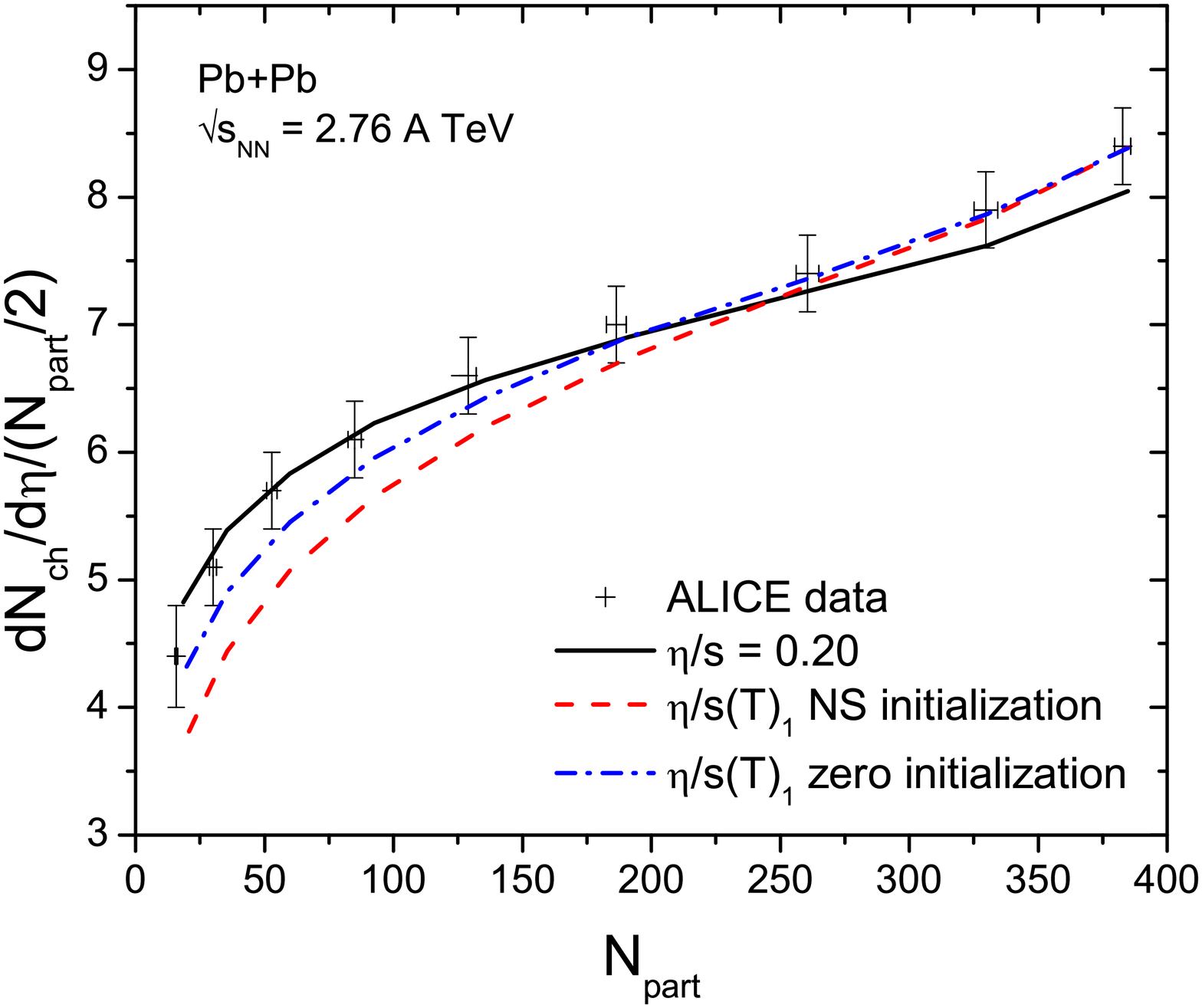}
\caption{(Color online) Final charged multiplicity per wounded nucleon 
pair as a function of number of participant nucleons in Pb+Pb collisions 
at $\sqrt{s}\eq2.76\,A$\,TeV, for different functional forms of
$(\eta/s)(T)$ and initial conditions for the
shear stress tensor
$\pi^{\mu\nu}$ (see text).
\label{F12} 
}
\end{figure}

Figure~\ref{F12} illustrates the influence of a linear temperature 
dependence of $\eta/s$ as in Eq.~(\ref{eq1}) on the centrality dependence 
of charged hadron production. The solid black line is the same as shown
in the upper part of Fig.~\ref{F1} where it forms the lower bound of
the shaded region; it corresponds to constant $\eta/s\eq0.2$ and 
Navier-Stokes initial conditions for the shear stress tensor, 
$\pi^{\mu\nu}\eq2\eta\sigma^{\mu\nu}$ at $\tau_0\eq0.6$\,fm/$c$. The 
dashed and dash-dotted lines in Fig.~\ref{F12} use $(\eta/s)_1(T)$
with either Navier-Stokes (dashed) or zero (dash-dotted) initial
conditions for $\pi^{\mu\nu}$. These last two lines were normalized
to the ALICE point for the $0{-}5\%$ most central Pb+Pb collisions
($\dNdeta\eq1584{\pm}80$ \cite{Aamodt:2010pb}), whereas the black line 
was normalized to our best guess before the ALICE data became available
($\dNdy\eq1800$, corresponding to $\dNdeta\eq1548$). The centrality
dependence is then controlled by the predictions from the MC-KLN model, 
modified by viscous entropy production during the hydrodynamic evolution.

We see that even a relatively modest temperature dependent increase
of $\eta/s$ in the QGP phase leads to a significantly stronger 
non-linearity in the dependence of charged particle production on
the number of wounded nucleons. The reason is that an increase of 
$\eta/s$ with temperature leads to more viscous heating in central
collisions (which probe higher initial temperatures and such larger
effective shear viscosities) than in peripheral ones (whose initial 
temperatures are lower). Since the entropy production rate is given by
\begin{equation}
  \partial_{\mu} S^{\mu} = \frac{\pi^{\mu\nu}\pi_{\mu\nu}}{2\eta T},
  \label{eq4}
\end{equation}
this effect is stronger for Navier-Stokes initial conditions (where 
$\pi^{\mu\nu}$ is proportional to the velocity shear tensor
$\sigma^{\mu\nu}$ which at early times diverges like $1/\tau$)
than for zero initial shear stress (where $\pi^{\mu\nu}$ starts from 
zero and approaches its Navier-Stokes value $2\eta\sigma^{\mu\nu}$ only
after several relaxation times $\tau_\pi$ when, due to its $1/\tau$ decay,
it has already decreased to much smaller values).\footnote{For reference
  we list the fractions of the finally measured entropy in the most
  central and most peripheral centrality bins shown in Fig.~\ref{F12}
  that are generated by viscous heating during the hydrodynamic expansion:
  Constant $\eta/s\eq0.2$: $\Delta S/S_\mathrm{final}\eq26\%$ ($0{-}5\%$)
  and 33\% ($70{-}80\%$); $(\eta/s)_1(T)$ with $\pi^{\mu\nu}_0\eq0$: 
  $\Delta S/S_\mathrm{final}\eq25\%$ ($0{-}5\%$) and 15\% ($70{-}80\%$); 
  $(\eta/s)_1(T)$ with $\pi^{\mu\nu}_0\eq2\eta\sigma^{\mu\nu}$: 
  $\Delta S/S_\mathrm{final}\eq60\%$ ($0{-}5\%$) and 49\% ($70{-}80\%$).}

If one were to postulate the validity of the MC-KLN model as the correct 
description of the initial particle production, the ALICE data shown in 
Fig.~\ref{F12} would exclude a temperature dependence of $\eta/s$ as given 
in Eqs.~(\ref{eq1}) and (\ref{eq2}) for Naver-Stokes initial conditions. 
While we are not prepared to make such a statement on the basis of 
Fig.~\ref{F12} alone, we believe that it is important to point out this
relatively strong sensitivity of the centrality dependence of $\dNdeta$ 
to the transport properties of the expanding fireball medium and to 
emphasize the constraints it thus places on possible models for the QGP 
shear viscosity. 

\begin{figure}[hbt]
\includegraphics[width=\linewidth,clip=]{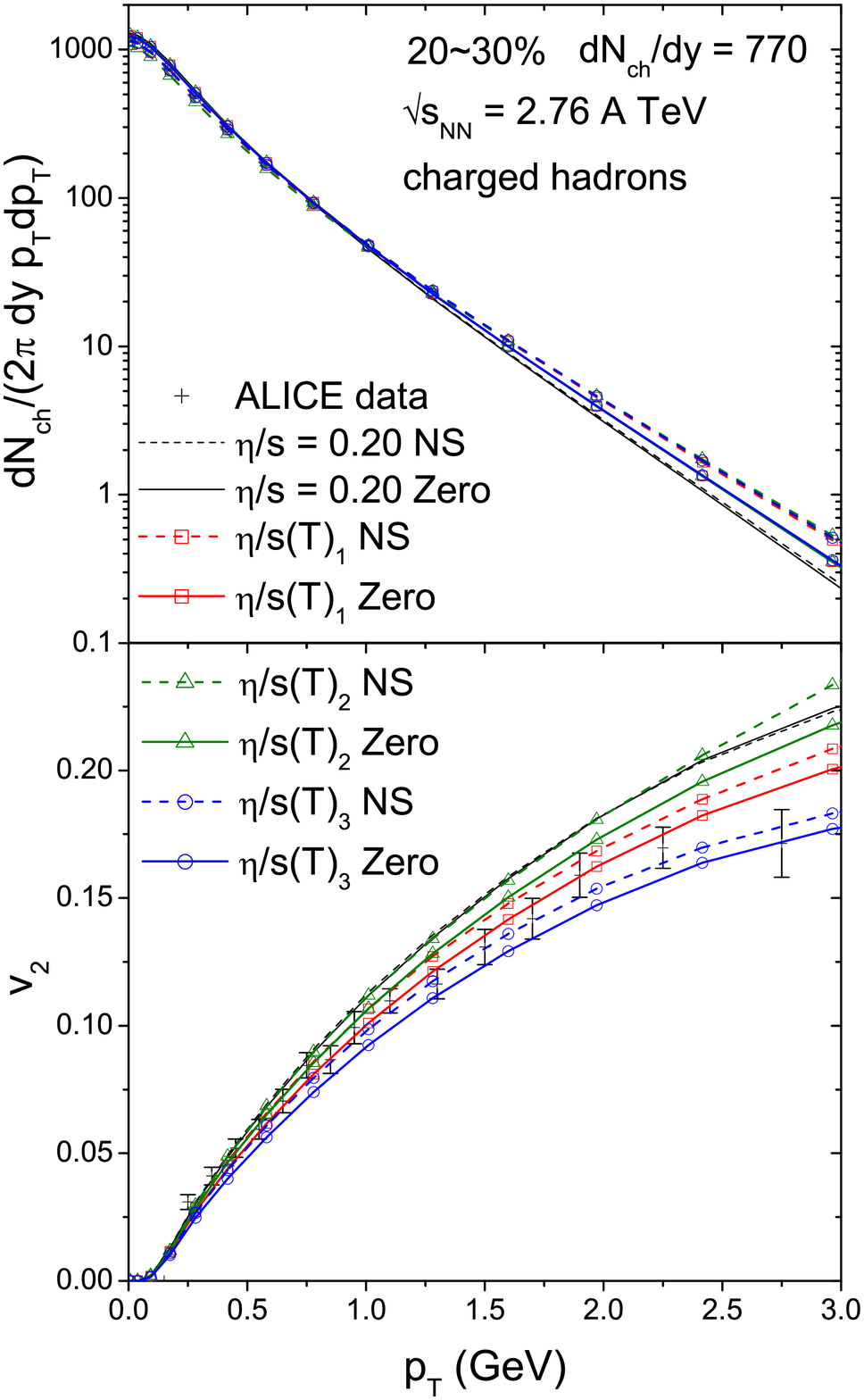}
\caption{(Color online)
Charged hadron transverse momentum spectra (top) and differential
elliptic flow (bottom) for $2.76\,A$\,TeV Pb+Pb collisions at $20{-}30\%$
centrality, for different models for the temperature dependence 
of $\eta/s$ and different initial conditions for $\pi^{\mu\nu}$
(Navier-Stokes (``NS'') or 0 (``Zero'')). The ALICE data in the bottom
panel are from Ref.~\cite{Aamodt:2010pa}.
\label{F13} 
}
\end{figure}

We now turn to the discussion of the influence of a possible temperature 
dependence of $\eta/s$ on the charged hadron $p_T$-spectra and elliptic 
flow. Figure~\ref{F13} shows LHC predictions for $2.76\,A$\,TeV Pb+Pb 
collisions of $20{-}30\%$ centrality. To ensure comparability of the
different cases studied in this figure we simply normalized the initial
entropy density profile such that we always obtain $\dNdy\eq770$, i.e.
the same value that we had obtained before for constant $\eta/s\eq0.2$ 
at this centrality. We first note that for constant $\eta/s\eq0.2$, we
don't observe any significant difference in the charged hadron spectra 
and elliptic flow between zero and Navier-Stokes initialization for
$\pi^{\mu\nu}$. Turning to the temperature-dependent parametrizations
$(\eta/s)_i(T)$, we note that for zero initialization of $\pi^{\mu\nu}$ 
(solid lines) our results agree with those reported in \cite{Niemi:2011ix}: 
An increase of $\eta/s$ at higher QGP temperatures leads to somewhat harder
charged hadron $p_T$-spectra (i.e. somewhat stronger radial flow, caused
by the larger tranverse effective pressure gradients at early times) and a 
suppression of the differential elliptic flow (due to an increase of
the time-averaged effective shear viscosity of the fluid). It is interesting 
to observe the hierarchy of the curves in Fig.~\ref{F13} corresponding
to the three parametrizations (\ref{eq1})--(\ref{eq3}): For the $p_T$-spectra,
all three $T$-dependent viscosities lead to almost identical hardening
effects on the spectral slope, while for the differential elliptic flow
$v_2^\mathrm{ch}(p_T)$ the curves are ordered not according to the
$\eta/s$-values at the initial central fireball temperature (see 
Table~\ref{T1}), but according to their hierarchy in the 
$165{\,<\,}T{\,<\,}280$\,MeV range. In fact, the observed magnitudes of 
the viscous $v_2$ suppression for the three $(\eta/s)(T)$ functions suggest 
that, at this beam energy and collision centrality, the buildup of elliptic 
flow is dominated by the QGP transport properties at 
$200{\,\alt\,}T {\,\alt\,}250$\,MeV. (At RHIC energies, the transport
properties for $T {\,\alt\,}200{-}220$\,MeV dominate the generation of
$v_2$ \cite{Niemi:2011ix}.)
 
%
\begin{table}[t!] 
\caption{Initial central entropy densities $s_0$ and temperatures $T_0$
  for the viscous hydrodynamic simulations of $20{-}30\%$ centrality 
  Pb+Pb collisions at the LHC ($\sqrt{s}\eq2.76\,A$\,TeV) shown in 
  Fig.~\ref{F13}. The different models for the $T$-dependence of $\eta/s$ 
  are defined in Eqs.~(\ref{eq1})--(\ref{eq3}). ``0'' stands for 
  $\pi_0^{\mu\nu}\eq0$ at $\tau_0$, ``NS'' stands for Navier-Stokes
  initialization of the shear stress tensor, 
  $\pi_0^{\mu\nu}\eq2\eta\sigma^{\mu\nu}$ at $\tau_0$.
  \label{T1}}
\begin{center}
\begin{ruledtabular}
\begin{tabular}{cccc}
  $\eta/s$ model &
  $\pi_0^{\mu\nu}$ &
  $s_0$\,(fm$^{-3}$) &  
  $T_0$\,(MeV) 
  \\
  \hline\hline
  $\eta/s\eq0.2$ &
  \begin{tabular}{c}
  0 \\ NS
  \end{tabular} 
  &
  \begin{tabular}{c}
  191.6 \\ 172.4
  \end{tabular} 
  &
  \begin{tabular}{c}
  427.9 \\ 413.9
  \end{tabular} 
  \\
  \hline
  $(\eta/s)_1(T)$ &
  \begin{tabular}{c}
  0 \\ NS
  \end{tabular} 
  &
  \begin{tabular}{c}
  179.6 \\ 119.3
  \end{tabular} 
  &
  \begin{tabular}{c}
  419.2 \\ 368.7
  \end{tabular} 
  \\
  \hline
  $(\eta/s)_2(T)$ &
  \begin{tabular}{c}
  0 \\ NS
  \end{tabular} 
  &
  \begin{tabular}{c}
  179.6 \\ 115.6
  \end{tabular} 
  &
  \begin{tabular}{c}
  419.2 \\ 365.1
  \end{tabular} 
  \\
  \hline
  $(\eta/s)_3(T)$ &
  \begin{tabular}{c}
  0 \\ NS
  \end{tabular} 
  &
  \begin{tabular}{c}
  175.2 \\ 116.6
  \end{tabular} 
  &
  \begin{tabular}{c}
  416.0 \\ 366.1
  \end{tabular} 
  \\
\end{tabular}
\end{ruledtabular}
\end{center}
\end{table}
%

For Navier-Stokes initial conditions (dashed lines in Fig.~\ref{F13}),
the increase in radial flow caused by an increase of $\eta/s$ at high
temperature is stronger and the viscous $v_2$ suppression is weaker than
for zero initial $\pi^{\mu\nu}$. This is caused by the much larger initial
shear stress tensor components in the NS case, compared to the case
of $\pi_0^{\mu\nu}\eq0$ where $\pi^{\mu\nu}$ approaches its (by that time
already much smaller) Navier-Stokes limit only after several relaxation 
times $\tau_\pi$ \cite{Song:2007fn}. The increase of $\eta/s$ with 
temperature generates a steeper initial transverse effective pressure gradient
(since $\pi^{\mu\nu}$ grows faster than the entropy density $s$ when
$\eta/s$ increases with temperature), and this generates stronger radial
flow. It also causes a larger spatial eccentricity of the initial 
effective pressure profile which (when compared to the case of 
$\pi_0^{\mu\nu}\eq0$) generates stronger elliptic flow. In fact, we found 
that for earlier starting times $\tau_0$ (where the Navier-Stokes values
for $\pi^{\mu\nu}$ are even larger), the quadratic parametrization 
$(\eta/s)_2(T)$ with NS initial conditions can lead to {\em more} elliptic
flow than a constant $\eta/s\eq0.2$, in spite of the larger mean viscosity 
of the fluid. 

We conclude from this exercise that a firm determination whether or not 
the ALICE data point towards a temperature-dependent growth of $\eta/s$
with increasing $T$, as expected from perturbative QCD \cite{Arnold:2003zc}
and (perhaps) from lattice QCD \cite{Meyer:2007ic}, is not possible without
a better understanding of the initial conditions for the energy momentum 
tensor (in particular the shear stress components) at the beginning
of the hydrodynamic evolution. Whereas generically larger viscosities
cause a suppression of the elliptic flow, temperature-dependent viscosities
can influence the initial effective pressure profile and its eccentricity
in a way that counteracts this tendency and, for some models such as
Navier-Stokes initial conditions, can even overcompensate it. 

\section{Conclusions}
\label{sec5}

Based on an successful global fit of soft hadron production data in 
$200\,A$\,GeV Au+Au collisions at RHIC with a pure viscous hydrodynamic 
model with Cooper-Frye freeze-out, presented in Sec.~\ref{sec2}, we 
generated hydrodynamic predictions for the $p_T$-spectra and differential
elliptic flow of unidentified charged hadrons and identified pions and 
protons for Pb+Pb collisions at the LHC. Where available, these predictions
were compared with available experimental data from the ALICE Collaboration.
Our extrapolation from RHIC to LHC energies was based on the assumption
that the QGP shear viscosity $\eta/s$ does not change with increasing fireball
temperature and stays fixed at the value $\eta/s\eq0.2$ extracted from the 
RHIC data, assuming MC-KLN initial conditions. The start time $\tau_0$ for 
the hydrodynamic evolution and the freeze-out temperature $\Tdec$ were held 
fixed, too. We found that, using the beam energy scaling implicit in the
MC-KLN model, such an extrapolation gives a good description of the centrality
dependence of charged hadron production and the charged hadron $p_T$-spectra
in central Pb-Pb collisions, but overpredicts the slope of the 
$p_T$-differential elliptic flow and the value of its $p_T$-integrated 
value by about $10{-}15\%$ in mid-central to mid-peripheral collisions.
In the most peripheral collisions, the predicted charged hadron $p_T$-spectra
are too flat, and the integrated elliptic flow is too small compared to
the experimental data. A preliminary study of possible temperature dependent
variations of $\eta/s$ in the high-temperature region explored for the first
time at the LHC remained unconclusive but pointed to a clear need
for better theoretical control over the initial conditions for the 
hydrodynamic energy-momentum tensor, in particular its
shear stress
components. The development of detailed dynamical models for the
pre-thermal evolution of the collision fireball and their matching to
the viscous hydrodynamic stage is a matter of priority for continued
progress towards quantifying  the transport properties of the quark-gluon 
plasma at different temperatures and densities. 


\vspace*{-5mm}

\acknowledgments{We would like to thank R. Snellings and A. Tang for 
providing us with tables of the experimental data from the ALICE 
experiment and for helpful discussions. This work was supported by the 
U.S.\ Department of Energy under contracts DE-AC02-05CH11231, 
DE-FG02-05ER41367, \rm{DE-SC0004286}, and (within the framework of 
the JET Collaboration) \rm{DE-SC0004104}. P.H.'s research was
supported by the ExtreMe Matter Institute (EMMI) and by BMBF under 
contract no.\ 06FY9092. 
}


\end{document}